\documentclass[	DIV=calc,%
							paper=a4,%
							fontsize=11pt,%
							twocolumn]{scrartcl}	 					

\usepackage{lipsum}													
\usepackage{mdwlist}
\usepackage[english]{babel}										
\usepackage[protrusion=true,expansion=true]{microtype}				
\usepackage{amsmath,amsfonts,amsthm}					
\usepackage[pdftex]{graphicx}									
\usepackage[svgnames]{xcolor}									
\usepackage[hang, small,labelfont=bf,up,textfont=it,up]{caption}	
\usepackage{epstopdf}												
\usepackage{subfig}													
\usepackage{booktabs}												
\usepackage{fix-cm}													
\usepackage{url}

\usepackage{sectsty}													
\allsectionsfont{
	\usefont{OT1}{phv}{b}{n}
	}

\sectionfont{
	\usefont{OT1}{phv}{b}{n}
	}

\usepackage{fancyhdr}												
\usepackage{lastpage}	

\usepackage{lettrine}
\newcommand{\initial}[1]{%
     \lettrine[lines=3,lhang=0.3,nindent=0em]{
     				\color{DarkGoldenrod}
     				{\textsf{#1}}}{}}

\usepackage{titling}															

\newcommand{\HorRule}{\color{DarkGoldenrod}
									  	\rule{\linewidth}{1pt}%
										}

\pretitle{\vspace{-30pt} \begin{flushleft} \HorRule 
				\fontsize{50}{50} \usefont{OT1}{phv}{b}{n} \color{DarkRed} \selectfont 
				}
\title{Timing Analysis of SSL/TLS Man in the Middle Attacks}					
\posttitle{\par\end{flushleft}\vskip 0.5em}

\preauthor{\begin{flushleft}
					\large \lineskip 0.5em \usefont{OT1}{phv}{b}{sl} \color{DarkRed}}
\author{Kevin Benton, Ty Bross }											
\postauthor{\footnotesize \usefont{OT1}{phv}{m}{sl} \color{Black} 
					INFO I-521:  Spring 2012 								
					\par\end{flushleft}\HorRule}

\date{}																				

\begin{document}
\maketitle
\thispagestyle{fancy} 			
\section*{Abstract}
\initial{M}\textbf{an in the middle attacks are a significant threat to modern e-commerce and online communications, even when such transactions are protected by TLS.  We intend to show that it is possible to detect man-in-the-middle attacks on SSL and TLS by detecting timing differences between a standard SSL session and an attack we created.}

\section*{Introduction}
A man-in-the-middle attack is one in which an adversary places himself between legitimate users, masquerading as each of the legitimate users when communicating with the other.  The SSL/TLS version of this is as follows:

\begin{figure}[h]
\centering
\includegraphics[width=7cm,height=3cm]{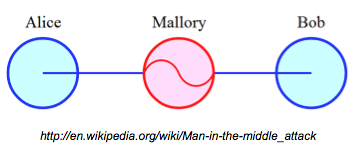}
\caption{Man in the Middle attack}
\label{fig:151106b_measure}
\end{figure}

\begin{enumerate*}
	\item Alice tries to start an SSL handshake with Bob but the connection is intercepted by Mallory 
	\item Mallory starts her own SSL session with Bob
	\item Mallory generates and sends a certificate that appears to be from Bob
	\item If Alice accepts the certificate, she establishes an SSL sesion using Mallory's key
	\item Messages from Alice are decrypted by Mallory and then sent over the SSL session she established with Bob
	\item Mallory can now relay messages between the SSL sessions while examining/modifying the plaintext data
\end{enumerate*}

The obvious issue above is that Alice accepted the certificate from Mallory that impersonated Bob. The public key infrastructure was designed to prevent this by having some trusted certificate authority authenticate and sign certificates to show that the subject in the certificate is the only entity that has the private key that goes with it. However, there are several issues (discussed below) that make the PKI much less trustworthy than originally intended. 

Our goal is to look at SSL man-in-the-middle attacks from a network traffic perspective to see if certain behaviors give away an attack. More specifically, we perform a timing analysis to determine if the attacker's certificate generation phase can be detected due to long response times after starting the SSL handshake.

\subsection*{Background}
The Stevens et al. demonstrated a collision attack that exploited the weakness of MD5 to generate an unauthorized valid intermediary CA certificate\cite{Stevens:2009:SCC:1615970.1615976}. They did this first by getting a regular server certificate from a trusted CA. Since the signature scheme was MD5, they were able to find a collision with a certificate they generated that had signing privileges. Since the signatures were the same MD5 hash, it effectively became an intermediary CA certificate signed by the original CA. This certificate could then be use to conduct man-in-the-middle attacks that would pass the PKI validation.

Moxie Marlinkspike introduced an interesting vulnerability in the way that browsers (used to) handle null termination characters that appear in certificates\cite{Marlinspike_2009}. He found that some certificate authorities would allow him to put a null character in the subject field of the certificate. Therefore, a request could be made for \emph{bank.com\\0.attacker.com} and the authority would correctly verify that he owned \emph{attacker.com} (since it's the parent domain) and issue the certificate. Then, due to browser implementations at the time, they would stop at the null char and accept the certificate as though it were for \emph{bank.com}. It even worked with wildcards, so an attacker could get a cert for  \emph{*\\0.attacker.com} and browsers would treat it as a certificate for \emph{*} (all domains), allowing it to be used to impersonate any website.

Current certificate stores trust many authorities and very few users actually check to see which authorities are trusted by their browser. There are various conditions where an additional CA could be added to the trust list without the user noticing\cite{canttrustssl}. This could be done by malicous adversaries via malware, or it could simply be done by the entity in charge of maintaining the computer. In either case, man-in-the-middle attacks can use the certificate to conduct "PKI-valid" man-in-the-middle attacks.

Like any other institutions, certificate authorities have to answer to the government of the country in which they operate. They could be required to just give a valid signing certificate to law enforcement agencies for wiretapping. Devices created specifically for this purpose were leaked to the public\cite{evilcops}, showing that these attacks occur often enough to warrant special hardware. Finally, certificate authorities can simply be hacked and have their certificates stolen, as was the case with DigiNotar\cite{wiki:diginotar}.

Even with the high theoretical security of the PKI, all of these attacks show that the current structure is vulnerable to political pressures and suffers from the problem where the weakest link breaks the whole system due to browsers trusting all authorities equally. Although certificate revokation mechanisms help with a lot of these problems, \cite{Marlinspike_2009} demonstrated that many browsers do not strictly enforce revokation checks so they can be blocked or responded to with an error code. This motivated us to look at other methods of detecting SSL MiTM attacks.

\section*{Experimental Design}
To measure the behavior of normal SSL handshakes compared to ones conducted with MiTM attackers, we wrote a program that  starts an SSL handshake and then closes the connection after receiving the certificate from the remote party. It measures the time it takes for the TCP connection to be established and the time it takes for the remote party to reply with a certificate. Due to the nature of certain MiTM attacks generating certificates in real-time after connecting to the real target, the gap between when the connection is established and when the certificate is received should widen significantly under the attack condition. 

Since the gap between the handshake hello and the server certificate includes network transport time, we subtract out an RTT to account for this. We roughly estimate the RTT by averaging the time taken for the remote server to respond to several TCP SYN packets. To account for jitter and other short-lived network conditions, the script collects 19 samples over a five minute period for each website to allow us to compare the averages between attack conditions.

We created a list of 40 domains from several distinct geographic locations, including the US, India, the United Kingdom, Europe, and Japan, that make use of SSL.  We then established a baseline for round-trip time (RTT) and certificate generation time for each of these sites by running the script without an MiTM attack. 

To analyze the behavior under MiTM attack conditions, we tested it against three popular software packages that have SSL man-in-the-middle capabilities.  These packages are \emph{WebMiTM}\cite{webmitm},  \emph{Cain \& Abel}\cite{cainabel}, and \emph{Ettercap}\cite{ettercap}.

\section*{Results}

\begin{figure*}[!htbp]
\centering
\includegraphics[height=8cm]{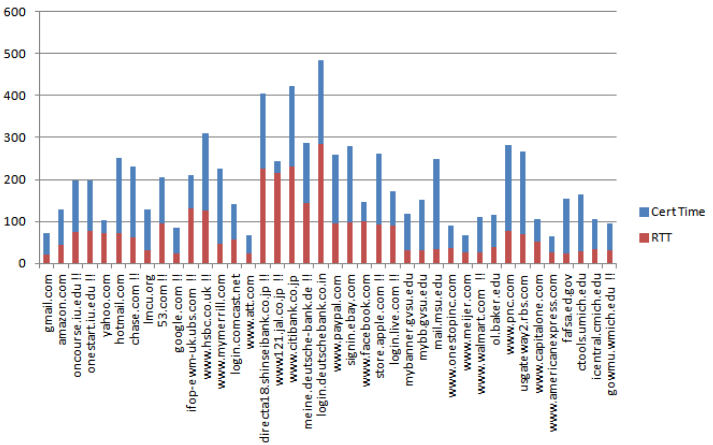}
\caption{Baseline Graph}
\label{fig:baseline}The average RTTs
\end{figure*}

Figure \ref{fig:baseline} shows the average RTT and certificate generation times for each of the 40 measured sites. "certificate generation times" in the context of this paper is the estimated time it takes the server send out the certificate from when it receives the \emph{ClientHello} message.

After examining the different timing profiles for each site, there didn't appear to be an observable trend across all of the websites. This is likely caused by the wide variety of web server software and how each handles certificate caching, connection queuing, and crypto computations in general. All of the sites we chose were high profile websites that could have a wide variation of loads. Additionally, it's possible for the higher security profile sites (e.g. banks) to be using an SSL proxy to protect their servers from attacks. Therefore, at this point we don't have a true negative identification method for SSL MiTM attacks.

\subsection*{Ettercap}
\begin{figure*}[!htbp]
\centering
\includegraphics[height=8cm]{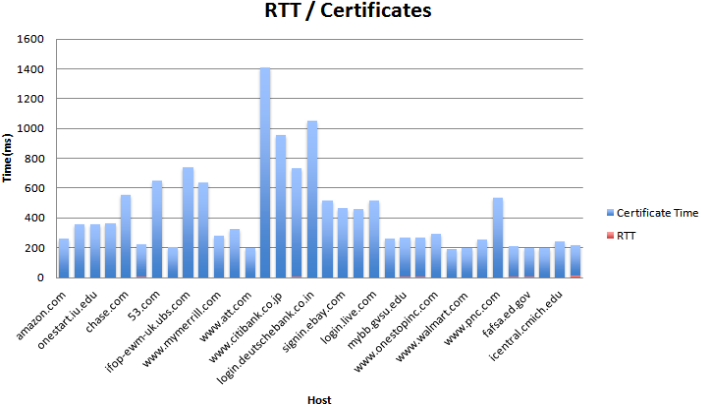}
\caption{Ettercap}
\label{fig:ettercap}
\end{figure*}

\begin{figure*}[!htbp]
\centering
\includegraphics[height=8cm]{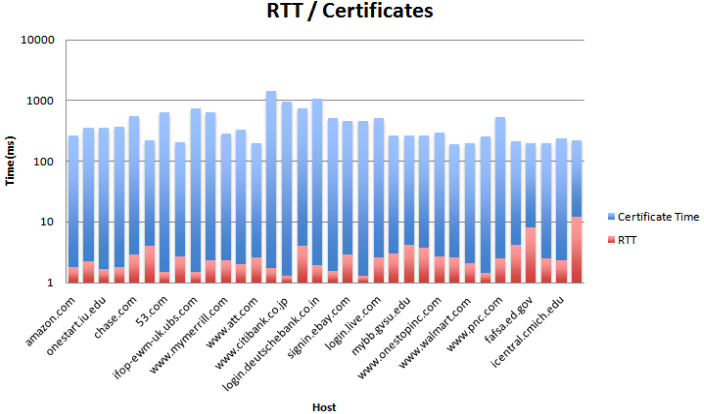}
\caption{Ettercap Logarithmic Scale}
\label{fig:ettercaplog}
\end{figure*}

Figure \ref{fig:ettercap} shows the average RTT and certificate times for the websites while the test server was victim to an \emph{Ettercap} ARP poisoning MiTM attack. Nearly all of the delay has been shifted from RTT to the certificate time. To illustrate it better, figure \ref{fig:ettercaplog} shows a logarithmic scale of the same chart. For all of the sites, the cert time was at least an order of magnitude more than the RTT time. 

We also observed that \emph{Ettercap} revealed itself in a non-timing related manner. It accept HTTPS connections and allow the \emph{ClientHello} before sending a RST when a client tries to connect to a non-HTTPS enabled website. So, a simple tool could detect \emph{Ettercap} MiTM attacks by attempting to establish a TCP connection to port 443 on a server that doesn't accept connections on that port, in which case Ettercap would accept the connection anyway.

\subsection*{WebMiTM}

\emph{WebMiTM} is different from \emph{Ettercap} and \emph{Cain \& Abel} in that it is only designed to handle DNS-based man-in-the-middle attacks. It expects there to be a rogue DNS server directing traffic to it. In DNS MiTM based attacks, there isn't a way to determine which server an incoming TCP connection had originally intended to connect to. \emph{WebMiTM} compensates for this by reading the server name identifier in the \emph{ClientHello}. A client could then test for its presence by sending sites that shouldn't be on the same server in the SNI field to see if the corresponding HTTP content comes back (i.e. a single SSL server shouldn't accept SSL connections and return the correct page for \emph{google.com} and \emph{bing.com}).

In any DNS-based SSL MiTM attacks, a client can detect catch-all attacks (i.e. ones that intercept all traffic) with the same method described for \emph{Ettercap}, since they have to accept SSL conections to get the SNI from the \emph{ClientHello}.

\begin{figure*}[!htbp]
\centering
\includegraphics[height=8cm]{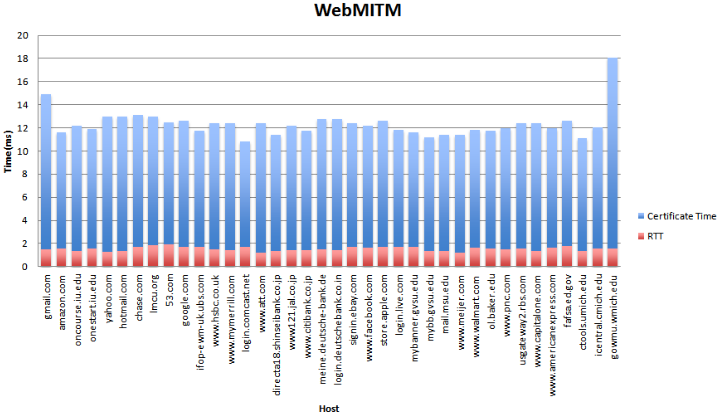}
\caption{WebMiTM}
\label{fig:webmitm}
\end{figure*}

From a timing perspective, \emph{WebMiTM} isn't very interesting (figure \ref{fig:webmitm}). There is almost no variance in the RTT or the cert time, but this is caused by a major limitation of \emph{WebMiTM}. It doesn't dynamically generate certificates to impersonate the correct website, instead it just uses one provided certificate for all of the connections. 

\subsection*{Cain \& Abel}
\emph{Cain \& Abel} performs ARP poisoning man-in-the-middle attacks on SSL sessions and can be configured to utilize a user-provided certificate to sign all of the certificates it generates for the web-servers that it impersonates. So, if an attacker obtains a trusted CA signing certificate and private key, it can be used here for "PKI-valid" attacks with a 'point-and-click' interface.

\begin{figure*}[!htbp]
\centering
\includegraphics[height=8cm]{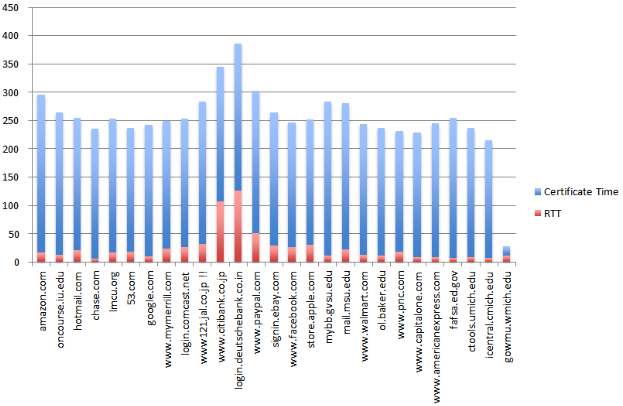}
\caption{Cain \& Abel RTT and Cert Time Averages}
\label{fig:cainNOTnormalized}
\end{figure*}

Figure \ref{fig:cainNOTnormalized} shows the average RTT and certificate generation times for 20 of the sites when \emph{Cain \& Abel} was used to conduct an ARP poisoning man-in-the-middle attack on the test machine. The average RTTs here are much lower; however, there was still a higher variance than we expected. 

This variance is caused by the way that \emph{Cain \& Abel} handles its certificate generation. It only appears to intercept a connection once it has listened to a normal handshake to extract the certificate and then generate and sign its own copy. Therefore, if a site hasn't been connected to before, the first SSL connection is allowed to go through using its normal certificate. This behavior was exacerbated by our program closing the connection during the middle of the handshake which seemed to interrupt \emph{Cain \& Abel's} process because it took it several connections to generate a certificate for some of the websites. 

\begin{figure*}[!htbp]
\centering
\includegraphics[height=8cm]{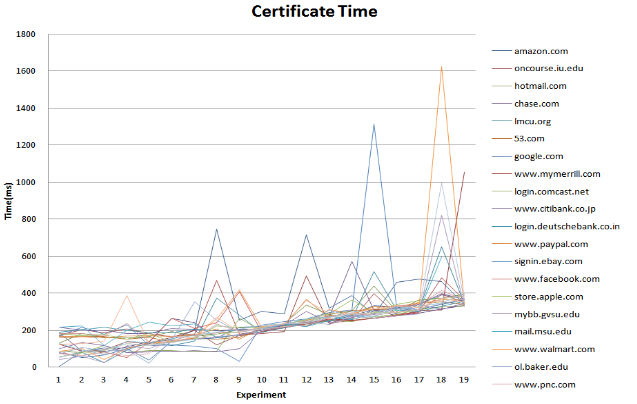}
\caption{Cain \& Abel Certificate Times for Individual Sites}
\label{fig:cainCERT}
\end{figure*}

\begin{figure*}[!htbp]
\centering
\includegraphics[height=8cm]{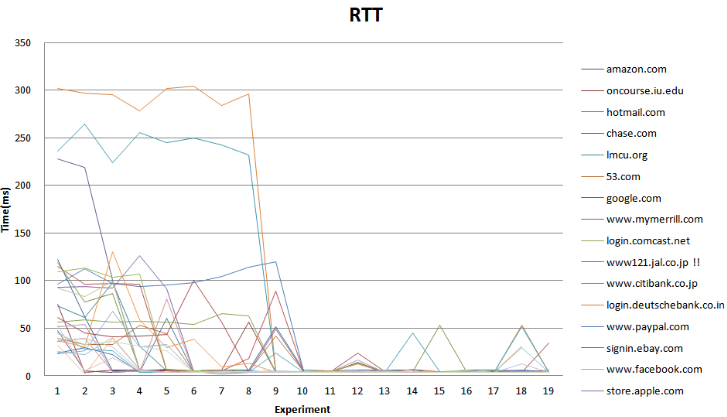}
\caption{Cain \& Abel RTT Times for Individual Sites}
\label{fig:cainRTT}
\end{figure*}

Figure \ref{fig:cainRTT} and figure \ref{fig:cainCERT} illustrate this behavior by showing the individual results from each test. The RTT for a site significantly declines  and its cert time increases once \emph{Cain \& Abel} successfully impersonates it. 

\begin{figure*}[!htbp]
\centering
\includegraphics[height=8cm]{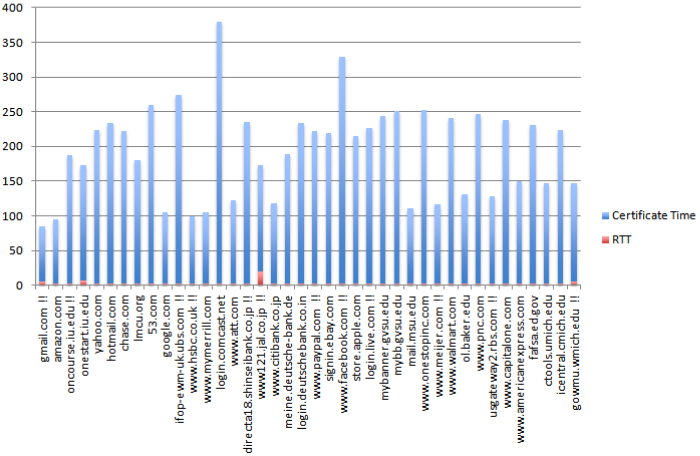}
\caption{Cain \& Abel RTT and Cert Time Averages (Only After Impersonation)}
\label{fig:cainnormalized}
\end{figure*}

We adjusted for this behavior by only measuring the times of the MiTM responses once \emph{Cain \& Abel} actually starts intercepting the SSL connections. Figure \ref{fig:cainnormalized} shows the times with this adjustment made. This matches the results received from \emph{Ettercap}, a low RTT time/variance between sites and a relatively high cert time/variance between sites.

An important distinction between \emph{Cain \& Abel} and the other two programs is how it handles connection requests. It directly mimics the server it's impersonating. This is important to consider when trying to be a completely transparent MiTM. For example, if a user knows that \emph{exmaple.org} does not listen for HTTPS connections on port 443, a connection attempt to port 443 should yield a TCP RST or a timeout. \emph{Cain \& Abel} is the only one that correctly does this.

The only way we found that \emph{Cain \& Abel} could be revealed (other than through timing analysis) is by monitoring certificate changes. Due to its passive behavior of allowing normal SSL connections until a certificate is generated, a user could connect to an arbitrary HTTPS site and then connect again a few seconds later to watch for fingerprint changes in the certificate. This method is still limited because there is the possibility that \emph{Cain \& Abel} could have already observed a connection to that same site and generated a certificate for it. 

\section*{Discussions/Future Work}
In our experiment, we illustrated timing differences inside an SSL handshake that occur during a man-in-the-middle attack. The key indicator in our results was the low RTT variance between websites. If the MiTM attacker is several hundred milliseconds away, the variance may increase to the point where it's no longer discriminating. To make our results more generalizable, future work would analyze various delays to the attacker and statistically define boundaries that identify an attack.

\section*{Related Work}
\subsection*{Detecting Attacks through Traffic Behavior}
Wagoner\cite{wagoner2011detecting} performed a similar analysis with the same goal of our project. However, his experimental design did not measure the internal timing of the handshake process leading to less generalizable results since just start-to-finish time was measured. The experiment was run on a local network where network delays were negligible, making any findings impossible to generalize to high latency variance networks like the Internet. For example, he observed an  increase from  \textless 10ms to $\approx$250ms with the MiTM tool he selected, which was the basis for his entire statistical analysis and we saw variations of up to 300ms just based on server location when measuring websites. 

DTRAB\cite{Fadlullah:2010:DCA:1959363.1959381} is a system that looks for anomolies in secure protocols to identify out-of-characteristic packets. It supports any protocol because it uses a machine learning approach to build profiles on inter-packet timing, payload sizes, flow sizes, etc. However, it wouldn't help against SSL MiTM attacks because the SSL flows would look the same as regular SSL traffic to their passive observation points. 

There are several other works that use timing analysis to link traffic to a source such as the research conducted by Wang et al. \cite{Wang:2002:IDB:646649.699363}, which correlates encrypted traffic relayed through multiple hops (e.g. SSH) to identify the source of traffic. Similarily, Abraham et al. \cite{abraham2010selective} presented timing attacks on low-latency mixing networks (e.g. Tor) to de-anonymize users.

To our knowledge, ours is the only work that examines the timing within an SSL/TLS handshake to detect SSL/TLS MiTM attacks. 

\subsection*{Additional Certificate Verification Mechanisms}
Another approach to detecting man-in-the-middle attacks is to perform additional verifications on the certificate received from the web server. These methods offer verifications that are difficult for a man-in-the-middle attack to circumvent.

Certlock \cite{Soghoian10certifiedlies:} employs a trust-of-first-use solution similar to the way SSH keys are handled. On connecting to an SSL site, the certificate is cached. If a different certificate is received on future connections, the certificate authority is checked against the cached copy. If the CA is from the same country, a warning is not issued. This criteria was chosen in response to sponsored eavesdropping by foreign countries. For example, an Iranian CA might be pressured by its government to issue a CA for gmail.com so they can read the email contents of network users. The issue with this approach is that an normal connection to the web server must be made to cache the certificate before it can detect attacks, which won't work in environments where MiTM attacks are persistent. 

Perspectives\cite{Wendlandt:2008:PIS:1404014.1404041}, SignatureCheck\cite{Jo:2011:SPD:2179298.2179365} and DoubleCheck\cite{conf/iscc/AlicherryK09} are methods of checking a received certificate using a side-channel. All of them use a PKI independent of the one trusted by the browser to establish a connection with third-party observation servers to request the thumbprints for a given web-site. Perspectives provides historically observed thumbprints from other clients, which results in false positives if a server recently changed its certificate. SignatureCheck connects to the web-server in question to get the thumbprints at the time of the request, which results in false negatives if the web-server in question is the victim of the man-in-the-middle attack rather than the client. DoubleCheck is similar to SignatureCheck, but it uses Tor for retrieval. 

\section*{Conclusion}
Our experiment revealed that SSL MiTM attacks have timing patterns that could be identified by observing victims. More specifically, the attack tools we analyzed shifted most of the delay to the time between when an SSL handshake was started and when the certificate was received. Additionally, most of the variance in the RTT was eliminated when connecting to sites all over the world due to the MiTM programs accepting TCP connections immediately. We also presented non-timing based methods to reveal when the specific MiTM tools we tested were being used. 

\bibliographystyle{abbrv}
\bibliography{paper-refs}  
\end{document}